%% file: paper.tex
\documentclass[conference]{IEEEtran}

\usepackage[draft]{minted}
\usepackage{cite}
\usepackage[pdftex]{graphicx}
\usepackage{subfig}
\usepackage[export]{adjustbox}
\usepackage{cellspace}
\usepackage{amsmath,amssymb,amsfonts}
\usepackage{algorithmic}
\usepackage{textcomp}
\usepackage{xcolor}
\usepackage{url}

\begin{document}

\title{Evaluating performance and portability of high-level programming models: Julia, Python/Numba, and Kokkos on exascale nodes}
\author{\IEEEauthorblockN{William F. Godoy, Pedro Valero-Lara, T. Elise Dettling, Christian Trefftz, Ian Jorquera, \\ Thomas Sheehy, Ross G. Miller,  Marc Gonzalez-Tallada, Jeffrey S. Vetter}
\IEEEauthorblockA{ 
Oak Ridge National Laboratory\thanks{This manuscript has been authored by UT-Battelle, LLC, under contract DE-AC05-00OR22725 with the US Department of Energy (DOE). The publisher acknowledges the US government license to provide public access under the DOE Public Access Plan (\protect\url{https://energy.gov/downloads/doe-public-access-plan}).}\\
\{godoywf\}, \{valerolarap\},  \{dettlingte\},  \{trefftzci\},  \{jorqueraid\},  \{sheehytb\}, \{rgmiller\},  \{gonzaleztalm\},  \{vetter\}@ornl.gov}

\IEEEauthorblockN{Valentin Churavy}
\IEEEauthorblockA{
Massachusetts Institute of Technology \\
vchuravy@mit.edu
}
}

\maketitle
\begin{abstract}
We explore the performance and portability of the high-level programming models: the LLVM-based Julia and Python/Numba, and Kokkos on high-performance computing (HPC) nodes: AMD Epyc CPUs and MI250X graphical processing units (GPUs) on Frontier's test bed Crusher system and Ampere's Arm-based CPUs and NVIDIA's A100 GPUs on the Wombat system at the Oak Ridge Leadership Computing Facilities.
We compare the default performance of a hand-rolled dense matrix multiplication algorithm on CPUs against vendor-compiled C/OpenMP implementations, and on each GPU against CUDA and HIP. Rather than focusing on the kernel optimization per-se, we select this naive approach to resemble exploratory work in science and as a lower-bound for performance to isolate the effect of each programming model.
Julia and Kokkos perform comparably with C/OpenMP on CPUs, while Julia implementations are competitive with CUDA and HIP on GPUs. Performance gaps are identified on NVIDIA A100 GPUs for Julia's single precision and Kokkos, and for Python/Numba in all scenarios. We also comment on half-precision support, productivity, performance portability metrics, and platform readiness.
We expect to contribute to the understanding and direction for high-level, high-productivity languages in HPC as the first-generation exascale systems are deployed.
\end{abstract}


\section{Introduction}
\input{01-introduction}

\section{Related Work}
\label{sec:Related Work}
\input{02-related}

\section{Experiments}
\label{sec:Experiments}
\input{03-experiments}

\section{Results}
\label{sec:Results}
\input{04-results}

\section{Performance Portability}
\label{sec:Portability}
\input{05-performance-portability.tex}

\section{Conclusions}
\label{sec:Conclusions}
\input{06-conclusions}

\section*{Acknowledgment}
This research was supported by the Exascale Computing Project (17-SC-20-SC), a collaborative effort of the US Department of Energy Office of Science and the National Nuclear Security Administration. This research used resources of the Oak Ridge Leadership Computing Facility at the Oak Ridge National Laboratory, which is supported by the Office of Science of the US Department of Energy under Contract No. DE-AC05-00OR22725.

\bibliographystyle{IEEEtran}
\bibliography{IEEEabrv,paper.bib}

\appendices

\section{Artifact Description for Reproducibility}
\label{ap1:Artifact}
\input{a1-appendix}

\end{document}

%% file: 01-introduction.tex
High-level dynamic languages such as  Python~\cite{van2007python}, Julia~\cite{Bezanson2017-ca}, and R~\cite{ihaka1996r} have been at the forefront of artificial intelligence/machine learning (AI/ML), data analysis, and interactive computing workflows in the last decade.
Traditional high-performance computing (HPC) frameworks that power the underlying low-level computations for performance and scalability are written in compiled languages: C, C\texttt{++}, and Fortran. At the same time, parallel programming models written in these languages aim to address the increasing heterogeneity of the targeted HPC hardware, which is dominated by highly multithreaded CPUs and graphics processing units (GPUs)~\cite{osti_1473756}.

The emergence and adoption of LLVM~\cite{lattner2004llvm} by major compiler vendors has led to unifying efforts to provide performance portable code across several languages and programming models. Julia and Python/Numba~\cite{lam2015numba} reuse LLVM's modularity by generating intermediate representations (LLVM-IR) to achieve performance, from their high-level, dynamic programming models. 
Similarly, directive-based standard approaches (e.g., OpenMP~\cite{openmp}, OpenACC~\cite{openacc}) provide a higher-level performance-portable model for HPC compiled languages, whereas metaprogramming approaches (e.g., Kokkos~\cite{Kokkos}, Raja~\cite{Raja,MarowkaHPCA22}, Thrust~\cite{Thrust}) provides powerful portable interfaces that target C\texttt{++} applications.
These high-level models rely on highly optimized vendor back ends (e.g., OpenMP, CUDA~\cite{cuda}, HIP~\cite{hip}), and their performance portability and overhead trade-offs have become an active area of research~\cite{MarowkaHPCA22}.
Nevertheless, high-level programming models become an attractive alternative to the end-to-end codesign process, thereby making them vital to closing gaps in the convergence of AI/ML, data science, and HPC as more emphasis is placed on the performance, portability, and productivity of scientific workflows~\cite{9309042}.

This work compares the performance, portability, and productivity of Julia, Python/Numba, and Kokkos high-level programming models for the CPU and GPU architectures that power upcoming exascale systems. We analyze single node scalability on two systems hosted at the Oak Ridge Leadership Computing Facility~(OLCF)\footnote{\url{https://www.olcf.ornl.gov/}}---Wombat, which uses Arm Ampere Neoverse CPUs and 2 NVIDIA A100 GPUs, and Crusher, which is equipped with AMD EPYC 7A53 CPUs and 8 MI250X GPUs and serves as a test bed for Frontier, the first exascale system on the TOP500 list.\footnote{\url{https://www.top500.org/}} We run hand-rolled general matrix multiplication~(GEMM) code for dense matrices using Julia, Python/Numba and Kokkos implementations and compare the performance with C for multithreaded CPU (OpenMP) and single GPU (CUDA/HIP) systems.
GEMM is an important kernel in the Basic Linear Algebra Subprograms (BLAS)~\cite{blackford2002updated} used across several deep learning AI frameworks, for which modern GPU architectures have been heavily optimized via tensor cores~\cite{7975270,7573804,8425458,9286214}.
The motivation for choosing a hand-rolled GEMM implementation is to i) isolate each programming model and environment in a simple kernel, and ii) to have a performance lower-bound point of reference that resembles custom scientific kernels in rapid prototyping formulations on dense matrices with many vector multiply and add operations.
Results are presented for implementations of half- (when possible), single-, and double-precision floating point operations. We evaluate the productivity and performance portability of these high-level approaches by using a common metric to analyze the resulting code implementations.

The rest of the paper is organized as follows: Section~\ref{sec:Related Work} provides a summary of related efforts that have evaluated the performance and portability of these high-level programming models. Section~\ref{sec:Experiments} describes the numerical experiments conducted on the Crusher and Wombat nodes. Performance results and follow-up discussion are presented in Section~\ref{sec:Results}, and the analysis of the performance portability is presented in Section~\ref{sec:Portability}. Section~\ref{sec:Conclusions} summarizes the study. Description of the reproducible artifacts used in this study are provided in Appendix~\ref{ap1:Artifact}.

%% file: 02-related.tex
Recent efforts have attempted to understand the performance gaps between portable high-level programming models and their equivalent highly optimized, vendor-specific implementations.
We classify this work according to the nature of the high-level implementation.

\paragraph{Dynamic Languages}  Julia provides a dynamic, compiled frontend to LLVM targeting scientific computing and data science. Its use in HPC is still an area of active exploration and community engagement~\cite{JuliaHPC}.
Ranocha et al.~\cite{Ranocha2022} present an assessment of their hyperbolic partial differential equation~(PDE) solver at scale, Trixi.jl. They conclude that although similar or even more complex challenges apply to Julia when running at scale, performance is similar to traditional HPC languages. Meanwhile, Tomasi and Giordano~\cite{Tomasi2018} explore the shortcomings and benefits of Julia for astrophysics applications.
Lin and McIntosh-Smith~\cite{9652798} use memory and compute-bound mini apps to show that Julia's performance is on par or slightly behind traditional compiled languages across several CPU/GPU HPC hardware configurations. 
Faingnaert et al.~\cite{9655458} provide optimized GEMM kernels in Julia that are competitive with cuBLAS and CUTLASS implementations.
Ko et al. introduce DistStat.jl~\cite{Ko2020}, which is a unified statistical computing environment in Julia for performance portability that has been tested on large-scale cloud systems. More recently, Giordano et al.~\cite{9912702} found competitive system performance for Julia's message passing interface (MPI)~\cite{snir1998mpi} MPI.jl~\cite{Byrne2021} on the Fujitsu A64FX Arm-based Fugaku system. Gmys et al.~\cite{Gmys2020} conclude that Julia and Python/Numba still present gaps for the scalability of multithreaded parallelizations when compared with Chapel~\cite{chamberlain2007parallel}.

Few recent efforts exist that leverage Python capabilities for performance via Numba. Mattson et al. present PyOMP~\cite{9658236}, which is an OpenMP implementation for Numba with preliminary results on par with C implementations that bypasses the Python's global interpreter lock (GIL). Recent studies on GPU implementations of Python/Numba target NVIDIA CUDA-supported hardware. For example, Oden~\cite{9092407} identifies gaps when comparing Numba's CUDA against C CUDA performance due to Python's performance limitations, whereas Di Domenico et al.~\cite{9756706} show promising performance when assessing NASA Advanced Supercomputing parallel benchmark kernels with Python. Python/Numba recently deprecated AMD GPU support,\footnote{\url{https://github.com/numba/numba/pull/6991}} whereas PyCUDA, PyOpenCL~\cite{KLOCKNER2012157}, and Cupy~\cite{nishino2017cupy} provide run-time access to NVIDIA and AMD GPU hardware by passing C or C\texttt{++} custom kernel code for compilation using a strings interface.

\paragraph{Meta-Programming} Meta-programming has become an intensive line of research for both code and performance portability. Kokkos~\cite{Kokkos}, Raja~\cite{Raja,MarowkaHPCA22}, and Thrust~\cite{Thrust} correspond to significant efforts in this area. They offer parallel dispatch options without specifying any detail of the target system.  Despite being used by many applications, Kokkos performance portability is still an active research subject~\cite{TeranishiDMB20,HalverMS20,abs-2103-11991,EllisR19}. Kokkos rely on highly optimized back ends (e.g., OpenMP, OpenACC~\cite{ValeroLaraLTDV22}, CUDA/HIP) that are based on template instantiations.  Thus hindering the deployment of kernel-specific optimizations (e.g., select the appropriate values for a number of blocks and threads per block, select the overlap of data transfers with computations). 
Templates set this kind of optimization, which happens earlier than the actual code generation phases. Every Kokkos' back end is an optimization of the common front end, it means that highly specialized techniques are used for parallel computation and memory management depending on the target back end or device (CPU or GPU). These may be different to the reference implementations used in this study, which can affect performance.

\paragraph{Directive-Based Languages} Directive-based languages are now ubiquitous within HPC. OpenMP and OpenACC are the de-facto standards for shared memory and accelerator programming. Code portability is addressed by compiler and run-time technology, which have proven sufficient to enable code porting across many HPC systems. For performance portability, however, this has not been the case. Both standards, and especially OpenMP, have increased their complexity by adding specialized constructs to guide the compiler in terms of what characteristics are available at the system level (e.g., unified virtual memory, vendor-specific features for the target device)~\cite{CatalanIWOMP20,Valero-LaraACCESS19,Valero-LaraJPDC20,Valero-LaraPDP19}.

\paragraph{Runtime Libraries} CUDA~\cite{cuda}, HIP~\cite{hip}, OpenCL~\cite{opencl}, and SyCL~\cite{sycl} have become common programming frameworks for HPC.
For instance, Bertoni et al.~\cite{BertoniIPDPSW20} studied performance portability of OpenCL on Intel and NVIDIA systems. Similarly, Halver et al.~\cite{HalverJS18} evaluated the portability of OpenCL for molecular dynamics applications. For SyCL, similar studies have been conducted to explore its portability to AI models~\cite{TanvirNIWOMCL22} and sparse linear algebra kernels~\cite{SabinoIWOMCL21}. In general, all these efforts suffer from some limitations, as vendor-specific run-time primitives are used by programmers to achieve high levels of performance. These primitives are not portable across HPC configurations, so they must be annotated with conditional compilation adding complexity to the whole process of deploying portable code. For performance portability, the same limitation arises and for the same reason---target-specific run-time primitives.

%% file: 03-experiments.tex
\begin{figure}[t]
    \centering
    \includegraphics[width=3.5in, height=1in]{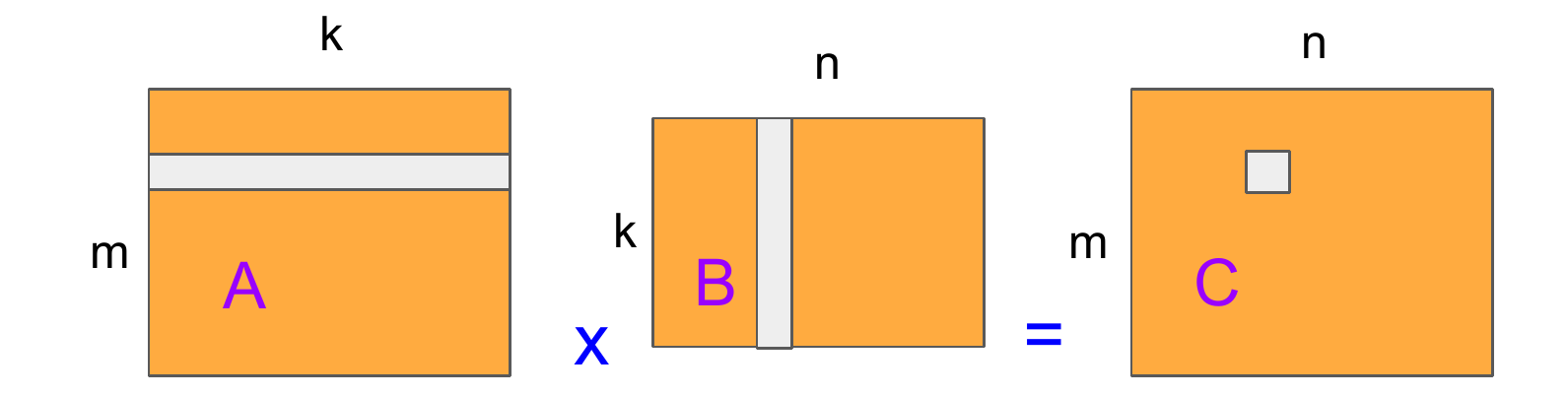}
    \caption{Schematic representation of a hand-rolled matrix multiplication for simple GEMM kernels.}
    \label{fig:gemm}
\end{figure}

This section describes different parallelization strategies on the targeted programming models for CPUs and GPUs. The hand-rolled GEMM kernel example is shown in Fig.~\ref{fig:gemm} as the product of two dense matrices.
Multithreaded CPU code implementations use coarse granularity mapping larger subcomponents per thread. These subcomponents are either entire rows or columns based on whether a language is row-major (e.g., Python default numpy arrays) or column-major (e.g., Julia) to ensure equivalent computational workloads.
Vectorized GPU code implementations use fine granularity mapping smaller subcomponents per thread. These subcomponents are singular elements defined by the 2D thread grid on GPU programming models. The rest of the section highlights the differences between programming models, compilation, and environment setup.

\subsection{CPU Implementations}

\begin{figure}[ht]

\centering

\begin{minipage}{1\columnwidth}
\centering (a) C/OpenMP 
\begin{minted}[breaklines, breakafter=d, fontsize=\scriptsize, frame=single, encoding=utf8]{c}
#include "omp.h"
...
#pragma omp parallel for default(shared) \ 
        private(i, k, j, temp)
for (i = 0; i < A_rows; i++)
  for (k = 0; k < A_cols; k++)
    temp = A[i * A_cols + k];
    for (j = 0; j < B_cols; j++) 
      C[i * B_cols + j] += temp * B[k * B_cols + j];
\end{minted}

\end{minipage}

\vspace*{1em}

\begin{minipage}{1\columnwidth}
\centering (b) Kokkos
\begin{minted}[breaklines, breakafter=d, fontsize=\scriptsize, frame=single, encoding=utf8]{c}
Kokkos::parallel_for( "AxB=C", mdrange_policy( {0, 0}, {M, N}),  KOKKOS_LAMBDA ( int m, int n ){
    float tmp = 0.0;
    for ( int k = 0; k < K; k++ )
      tmp += A(m, k) * B(k, n);
    C(m, n) = tmp;
  }
);
\end{minted}

\end{minipage}

\vspace*{1em}
    
\begin{minipage}{1\columnwidth}
\centering (c) Julia
\begin{minted}[breaklines, breakafter=d, fontsize=\scriptsize, frame=single, encoding=utf8]{julia}
import Base.Threads: @threads

function gemm(A, B, C)
...
    @threads for j in 1:B_cols
        for l in 1:A_cols
            @inbounds temp = B[l, j]
            for i in 1:A_rows
                @inbounds C[i, j] += temp * A[i, l]
            end
        end
    end
end
\end{minted}

\end{minipage}
\vspace*{1em}

\begin{minipage}{1\columnwidth}
\centering (d) Python/Numba
\begin{minted}[breaklines, breakafter=d, fontsize=\scriptsize, frame=single, encoding=utf8]{python}
from numba import njit, prange
import numpy as np

@njit(parallel=True, nogil=True, fastmath=True)
def gemm(A: np.ndarray, B: np.ndarray, C: np.ndarray):
...
    for i in prange(0, A_rows):
        for k in range(0, A_cols):
            temp = A[i, k]
            for j in range(0, B_cols):
                C[i, j] += temp * B[k, j]
\end{minted}
\end{minipage}

\caption{CPU multithreaded, coarse-granularity simple GEMM kernels for the programming models used in this study.}
\label{fig:GEMM-CPU}
\end{figure}

Coarse granularity for multithreaded CPU parallelization in C/OpenMP, Julia, and Python/Numba follows a similar approach to using metaprogramming directives on top of a serial \textit{for loop}-based implementation. On the other hand, Kokkos requires an anonymous lambda function implementation written entirely from the ground up. 
Figure~\ref{fig:GEMM-CPU} shows a typical single-level parallel \textit{for loop} implemented in C/OpenMP that provides pragmas with minimal modifications to a serial code version. As expected, index linearization of the multidimensional matrix is tracked by the programmer using non-safe memory access, whereas thread-private variables must be annotated to allow the compiler to find better optimizations. The number of threads is controlled with the \texttt{OMP\_NUM\_THREADS} environment variable. Additional thread policy is controlled by pinning the threads with the \texttt{OMP\_PROC\_BIND=true} and \texttt{OMP\_PLACES=threads} environment variables, as shown in Appendix~\ref{ap1:Artifact}.

Figure~\ref{fig:GEMM-CPU}b illustrates Kokkos's programming model that uses a C\texttt{++} lambda function to specify the calculations for an entry in the resulting matrix. Kokkos  aims  to  be  architecture agnostic  to  enable  programmers  to  move  past  the  low-level details  of  vendor-  or  target-specific  programming  models through  template specialization. 
In practice, the target architecture is defined at compilation time with the \texttt{KOKKOS\_DEVICES} flag.  
For  instance,  one  must  use \texttt{KOKKOS\_DEVICES=Cuda} to generate binary code for NVIDIA GPUs. The  Kokkos  library  and  the  source  code  are  then compiled for the targeted architecture.

The equivalent Julia implementation is shown in Fig.~\ref{fig:GEMM-CPU}c. Julia provides an even higher-level implementation that uses the built-in \texttt{Threads} module. As shown, the outer loop is immediately parallelized with the addition of the \texttt{@threads} macro without further specification. The downside of this approach is that the number of threads in Julia is immutable through an entire executable run because it is configured via a parameter, \texttt{-t}\,, to the Julia executable or the \texttt{JULIA\_NUM\_THREADS} environment variable. Owing to its numerical nature, Julia supports native multidimensional arrays and strong typing that can interoperate with the underlying multithreading back end implementation. The \texttt{@inbounds} macros prevent additional bound checks for array access. This can be configured at the executable level, but it is left in the code for illustration purposes. The \texttt{JULIA\_EXCLUSIVE} environment variable is used to control thread policy in all runs and pin threads to cores in strict order.

As shown in Fig.~\ref{fig:GEMM-CPU}d, Python/Numba provides a similar but slightly more invasive approach that uses metaprogramming decorators to mark the JIT compilation regions. The intended parallel for loop must be modified with the \texttt{prange} keyword. Although Numba supports numpy arrays, strong typing is not required within the JIT region. The \texttt{NUMBA\_NUM\_THREADS} environment variable allows one to select the number of threads, but there is currently no mechanism for setting a thread binding/pinning policy (unlike C/OpenMP and Julia).

The hand-rolled GEMM kernels shown in Fig.~\ref{fig:GEMM-CPU} were executed on two different CPU systems available at the OLCF: Wombat (Arm~+~NVIDIA) and Crusher (AMD). Table~\ref{tab:CPU} lists the CPU specifications, the required C/OpenMP and Kokkos compilation flags, the latest Julia and Python/Numba versions, and the environment variables (i.e., \texttt{ENV}). 
We selected target-specific flags on the corresponding LLVM-based, vendor-provided compilers (e.g., ArmClang, AMDClang) to ensure maximum on-node performance by using all available cores for a range of system workloads as defined by the matrix size. Overall, Julia and Python/Numba follow a similar approach to OpenMP for multithreaded codes, whereas Kokkos creates a portable unified API for both, the CPU and GPU.

\begin{table}[ht]
\centering
\caption{CPU experiment specs.}
\begin{tabular}{||p{0.3\linewidth} | p{0.26\linewidth}  | p{0.26\linewidth}||}
\hline
 Programming/System & Wombat (Arm)  &  Crusher (AMD)  \\
 Model & Ampere Altra & AMD Epyc 7A53 \\
  & 80-core, 1-NUMA      & 64-core, 4-NUMA \\
 \hline \hline 
 \textbf{C OpenMP} &  & \\
 Compiler & ArmClang22 & AMDClang14  \\
 Flags    & -O3 -fopenmp  & -O3 -fopenmp \\
          &               & -march=native \\
 \hline
 \textbf{C++ Kokkos}   & \multicolumn{2}{c||}{v3.6.01}   \\
 KOKKOS\_DEVICES &   \multicolumn{2}{c||}{OpenMP} \\
 KOKKOS\_ARCH & Armv8-TX2 & Zen~3 \\
 Compiler & ArmClang++22 & AMDClang++14 \\
 Flags    & -O3 -fopenmp & -O3 -fopenmp \\
          &              & -march=native \\
 \hline
 \textbf{Julia}  & v1.7.2 &  v1.8.0-rc1 \\
 ENV  & \multicolumn{2}{c||}{JULIA\_EXCLUSIVE=1} \\
 \hline
 \textbf{Python} & v3.9.9  &  \\
 \textbf{Numba}  &  v0.55.1 &  \\
 ENV  & \multicolumn{2}{c||}{NUMBA\_OPT=3  (default)}  \\ 
\hline
\end{tabular}
\label{tab:CPU}
\end{table}

\subsection{GPU Implementations}

GPU implementations follow a fine granularity approach by mapping the computations required to calculate an element of the resulting matrix to a single thread. The simple matrix multiplication kernel is described in Fig.~\ref{fig:GEMM-GPU}a for CUDA and HIP. HIP closely follows the CUDA kernel model, although the grid definition is based on the total number of launched threads, not blocks. 

\begin{figure}[ht]
\centering 
\begin{minipage}{1\columnwidth}
\centering (a) CUDA/HIP
\begin{minted}[breaklines, breakafter=d, fontsize=\scriptsize, frame=single, encoding=utf8]{c}
int row = blockIdx.y * blockDim.y + threadIdx.y;
int col = blockIdx.x * blockDim.x + threadIdx.x;
double sum = 0.0;
if( row < A_rows && col < B_cols )
{
  for(int i = 0; i < n; i++) {
    sum += A[row * n + i] * B[i * k + col];
  }
  C[row * k + col] = sum;
}
\end{minted}

\end{minipage}

\vspace*{1em}

\begin{minipage}{1\columnwidth}
\centering (b) Julia CUDA.jl
\begin{minted}[breaklines, breakafter=d, fontsize=\scriptsize, frame=single, encoding=utf8]{julia}
using CUDA
...
row = (blockIdx().x - 1) * blockDim().x + threadIdx().x
col = (blockIdx().y - 1) * blockDim().y + threadIdx().y

sum = zero(eltype(C))
if row <= size(A, 1) && col <= size(B, 2)
    for i in 1:size(A, 2)
        @inbounds sum += A[row, i] * B[i, col]
    end
    @inbounds C[row, col] = sum
end
return nothing

\end{minted}

\end{minipage}

\vspace*{1em}

\begin{minipage}{1\columnwidth}
\centering (c) Julia AMDGPU.jl
\begin{minted}[breaklines, breakafter=d, fontsize=\scriptsize, frame=single, encoding=utf8]{julia}
using AMDGPU
...
row = (workgroupIdx().x - 1) * 
       workgroupDim().x + workitemIdx().x
col = (workgroupIdx().y - 1) * 
      workgroupDim().y + workitemIdx().y
      
sum = zero(eltype(C))
if row <= size(A, 1) && col <= size(B, 2)
    for i in 1:size(A, 2)
        @inbounds sum += A[row, i] * B[i, col]
    end
    @inbounds C[row, col] = sum
end

\end{minted}
\end{minipage}

\vspace*{1em}

\begin{minipage}{1\columnwidth}
\centering (d) Python/Numba CUDA
\begin{minted}[breaklines, breakafter=d, fontsize=\scriptsize, frame=single, encoding=utf8]{python}
from numba import cuda
from numba.cuda.cudadrv.devicearray import DeviceNDArray
import numpy as np

@cuda.jit
def gemm(A: DeviceNDArray, B: DeviceNDArray, C: DeviceNDArray):
...
    i, j = cuda.grid(2)
    if i < C.shape[0] and j < C.shape[1]:
        tmp = 0.
        for k in range(A.shape[1]):
            tmp += A[i, k] * B[k, j]
        C[i, j] = tmp
\end{minted}

\end{minipage}

\caption{GPU fine-granularity, hand-rolled GEMM kernels for the programming models used in this study.}
\label{fig:GEMM-GPU}
\end{figure}

Julia GPU programming models use the vendor-specific CUDA.jl~\cite{besard2018juliagpu,besard2019prototyping} and AMDGPU.jl~\cite{AMDGPU} implementations for NVIDIA and AMD GPUs, respectively. They provide high-level mechanics to define multidimensional arrays (CUArray and ROCArray) on GPU devices. Julia also provides the KernelAbstractions.jl~\cite{KA} package for writing portable kernels while still maintaining dependence on either CUArray or ROCArray. Figures~\ref{fig:GEMM-GPU}b and~\ref{fig:GEMM-GPU}c show the corresponding kernel implementations on CUDA.jl and AMDGPU.jl, respectively. As shown, the close resemblance to the CUDA/HIP models for thread mapping makes for an easy transition for those familiar with these programming models. To its advantage, Julia uses multidimensional arrays and added functionality in the device kernel code to provide a powerful syntax for GPU programming.  

Python/Numba provides a very simple interface to access CUDA kernel functionality, as illustrated in Fig.~\ref{fig:GEMM-GPU}d. Unlike the CUDA/HIP model, it provides a simple \texttt{cuda.grid} mapping function between the row and column coordinate with a GPU thread. Similar to Julia, Python enables multidimensional matrix syntax on device kernels through the \texttt{devicearray} interface. As mentioned in Section~\ref{sec:Related Work}, Python/Numba support for AMD GPUs is currently deprecated.

\begin{table}[ht]
\centering
\caption{GPU experiment specs.}
\begin{tabular}{||p{0.3\linewidth} | p{0.25\linewidth} | p{0.31\linewidth}||}
\hline
Programming/System  & Wombat (NVIDIA)  &  Crusher (AMD)  \\
Model  & A100 Ampere & MI250X \\
 \hline \hline
 \textbf{C CUDA/HIP} & & \\
 Compiler & nvcc v11.5.1 & hipcc v14.0.0  \\
 Flags    & -arch=sm\_80 & --amdgpu-target=gfx908 \\
 \hline
 \textbf{C++ Kokkos}   & \multicolumn{2}{c||}{v3.6.01}   \\
 KOKKOS\_DEVICES & Cuda & Hip \\
 KOKKOS\_ARCH  & Ampere80 & Vega908 \\
 Compiler & CUDA v11.5.1 & HIP v14.0.0 \\
 Flags    & -expt-extended-lambda & --amdgpu-target=gfx908 \\
          & -Xcudafe & \\
          & -arch=sm\_80 & \\
 \hline
 \textbf{Julia}  & v1.7.2 &  1.8.0-rc1 \\
                 & CUDA.jl & AMDGPU.jl \\
                 &         &  \\
 \hline
 \textbf{Python} & v3.9.9  &  \\
 \textbf{Numba}  &  v0.55.1 &  Not supported \\
 Flags  &  & \\
\hline
\end{tabular}
\label{tab:GPU}
\end{table}

%% file: 04-results.tex
This section characterizes the results obtained in the experiments described in Section~\ref{sec:Experiments}. All numbers were obtained by running the GEMM kernels several (at least 5 or 10) times and excluding an initial warm-up step. This exclusion also discards initial communication (threads and GPUs) and JIT compilation overheads in Julia and Python/Numba. Due to the dedicated nature of the nodes, the results are the most likely performance value without doing an exhaustive variability analysis and only presenting the average expected value. We consider that variability is at face value a characteristic of the system, rather than an effect of the programming model per-se as it is the goal of this comparison. Nevertheless, reproducible artifacts are provided in Appendix~\ref{ap1:Artifact} for independent verification. 

\subsection{CPU Performance} 

\begin{figure}[ht]
    \centering
    \subfloat[Double precision (FP64)]{\includegraphics[width=3.5in,height=2.2in,valign=t]{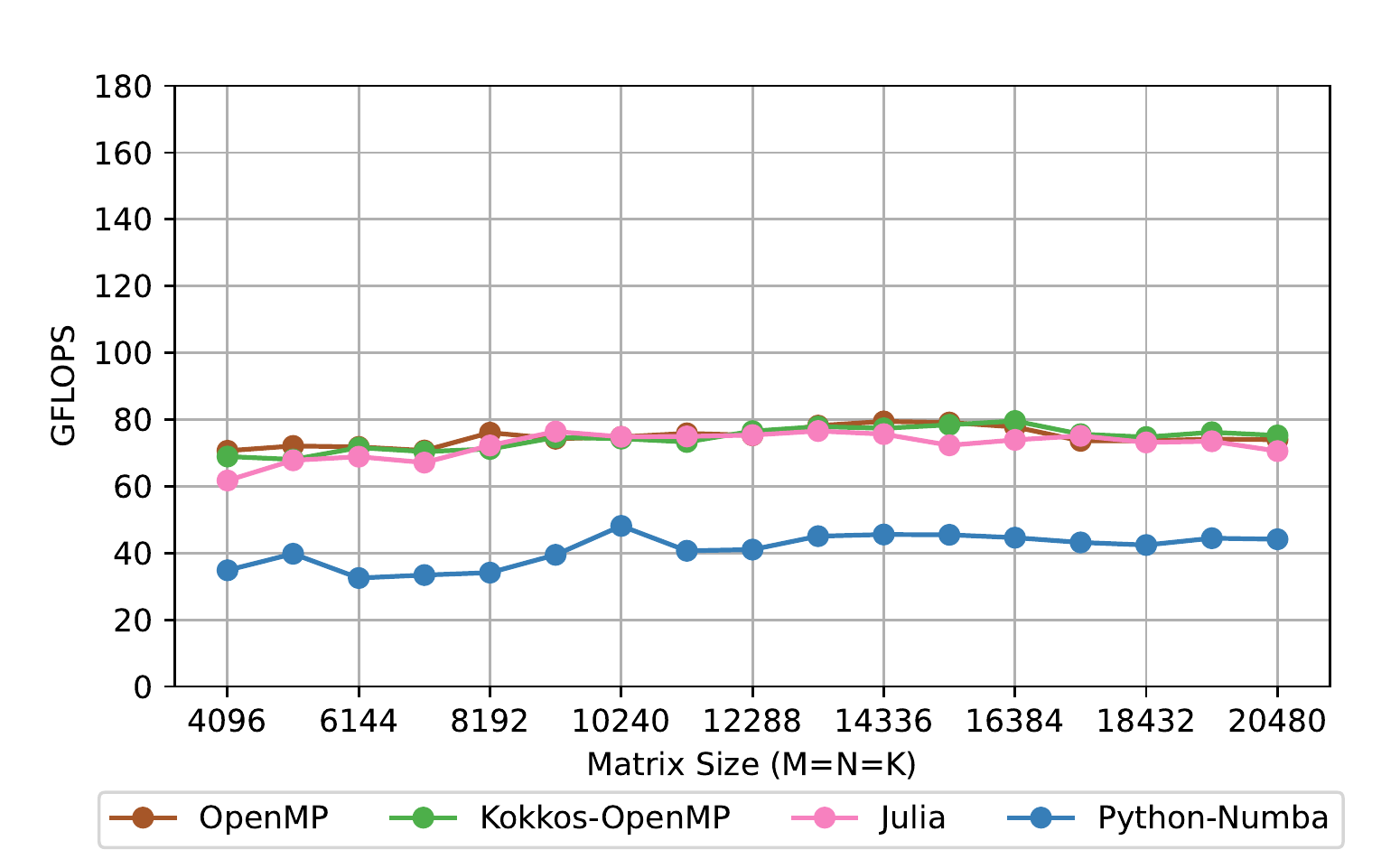}}
    \vspace{-0.6cm}
    \subfloat[Single precision (FP32)]{\includegraphics[width=3.5in,height=2.2in,valign=t]{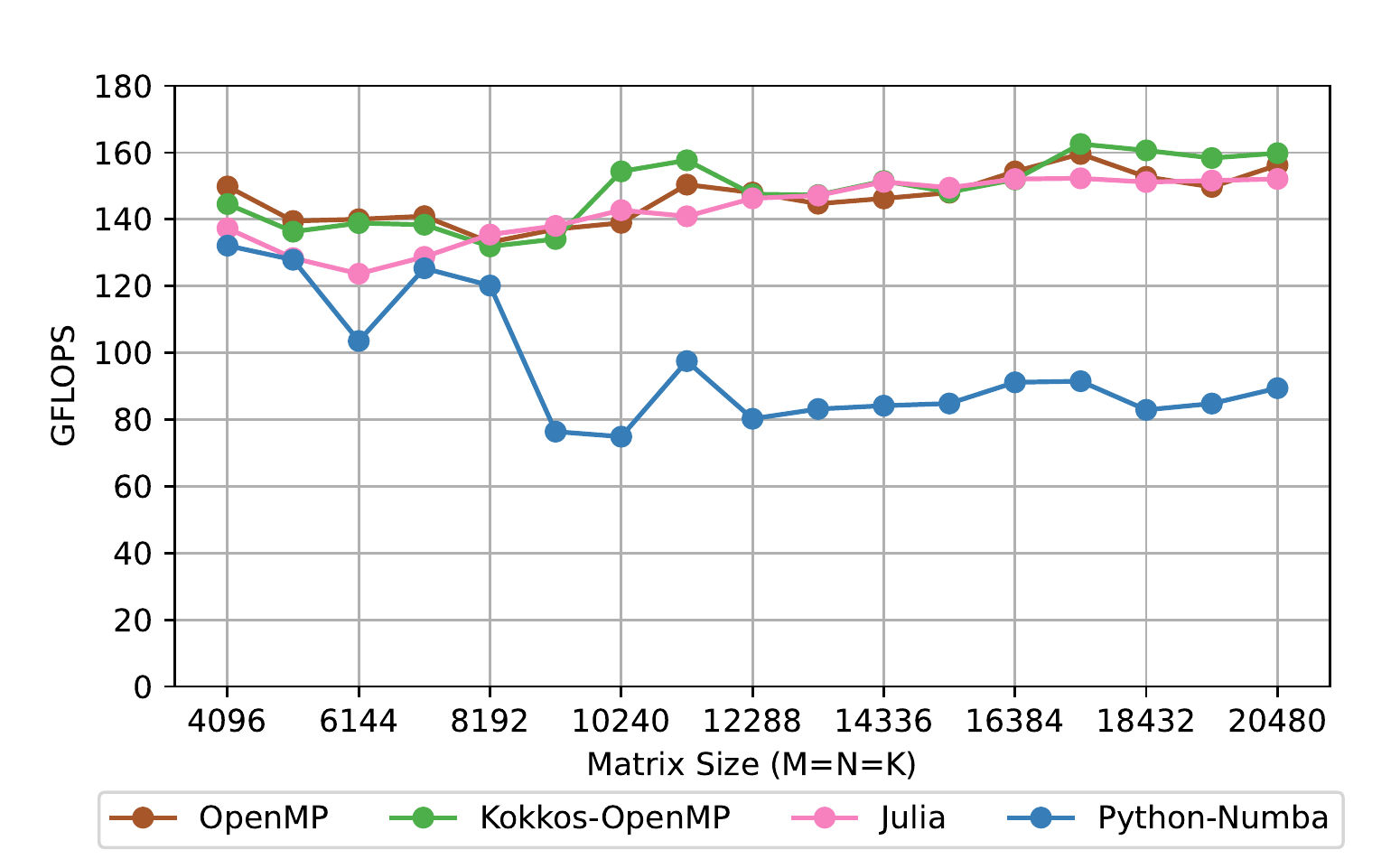}}
\caption{Crusher multithreaded CPU performance using 64 threads across 4 NUMA regions.}
\label{fig:CrusherAMDCPU}
\end{figure}

\paragraph{Crusher AMD EPYC 7A53} Figure~\ref{fig:CrusherAMDCPU} shows the results obtained for the multithreaded CPU implementations on the Crusher system for (a) double precision and (b) single precision. Overall, Kokkos/OpenMP and Julia threads perform comparably with the vendor ArmClang C/OpenMP implementation, whereas Python/Numba is still behind in terms of performance. OpenMP and Julia use environment flags to bind threads to CPU resources, as shown in Table~\ref{tab:CPU}; this option is not available in the Python/Numba APIs.

\begin{figure}[ht]
    \centering
    \subfloat[Double precision (FP64)]{\includegraphics[width=3.5in,height=2.2in,valign=t]{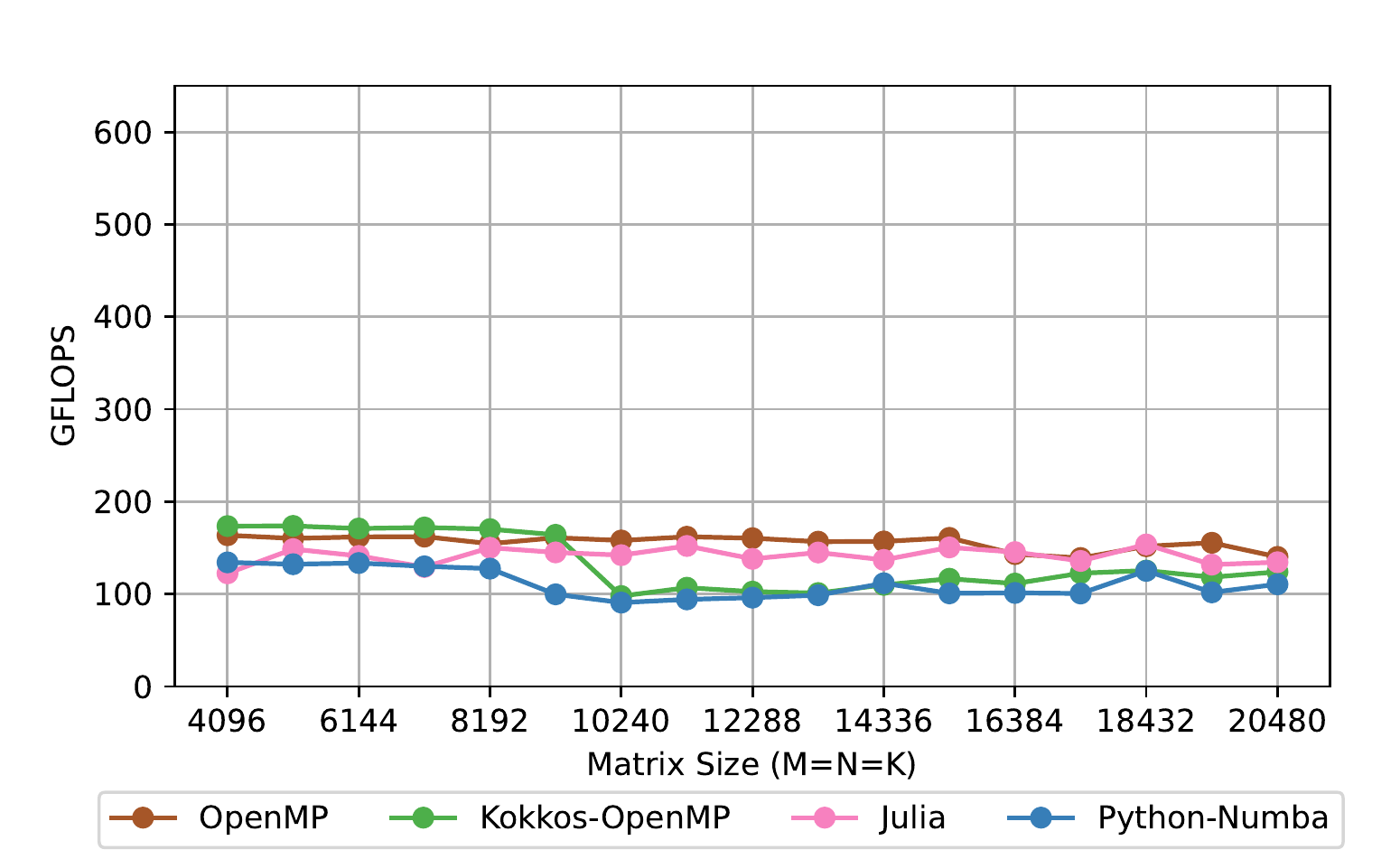}}
    \vspace{-0.6cm}
    \subfloat[Single precision (FP32)]{\includegraphics[width=3.5in,height=2.2in,valign=t]{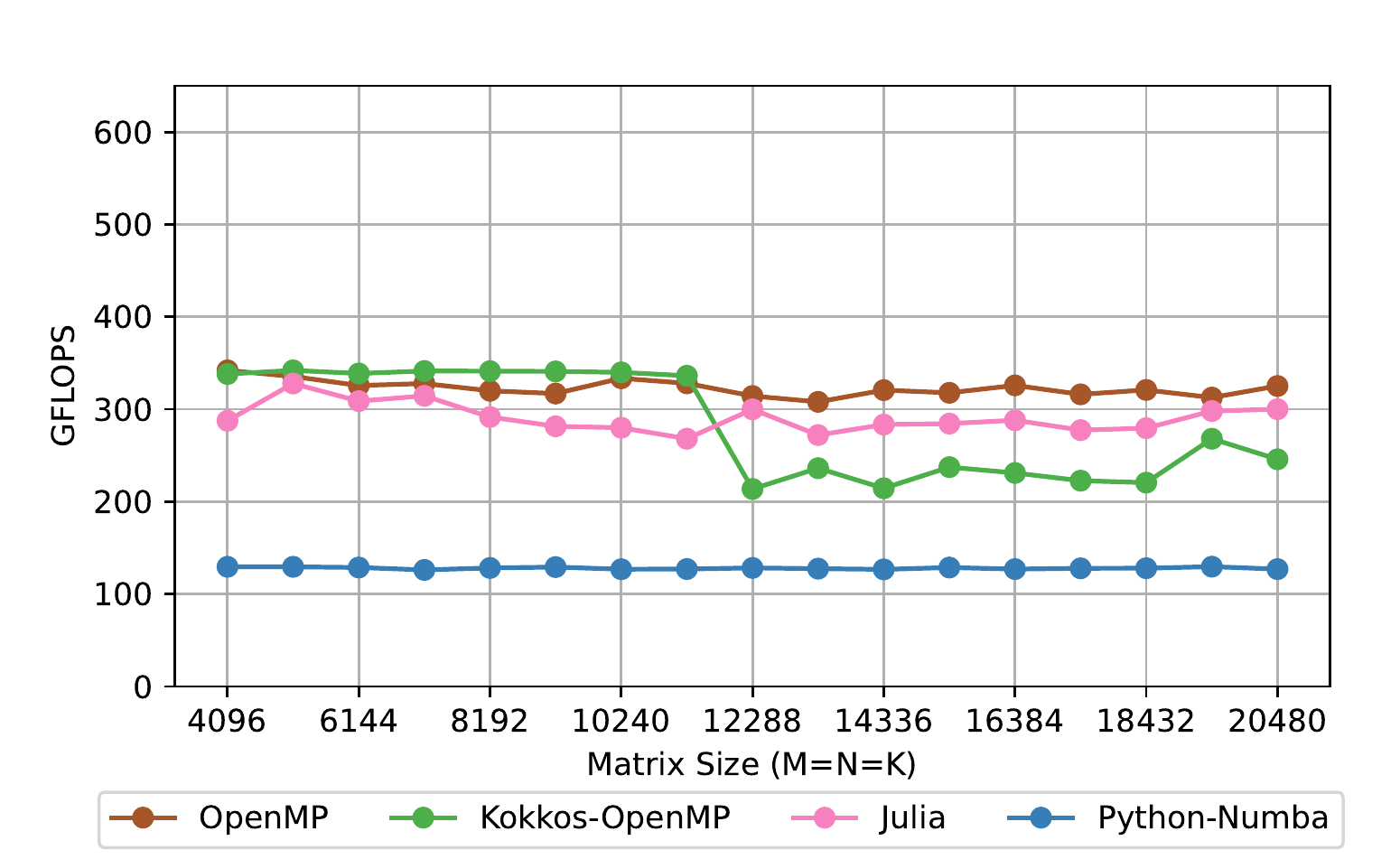}}
    \vspace{-0.6cm}
    \subfloat[Half precision (FP16)]{\includegraphics[width=3.5in,height=2.2in,valign=t]{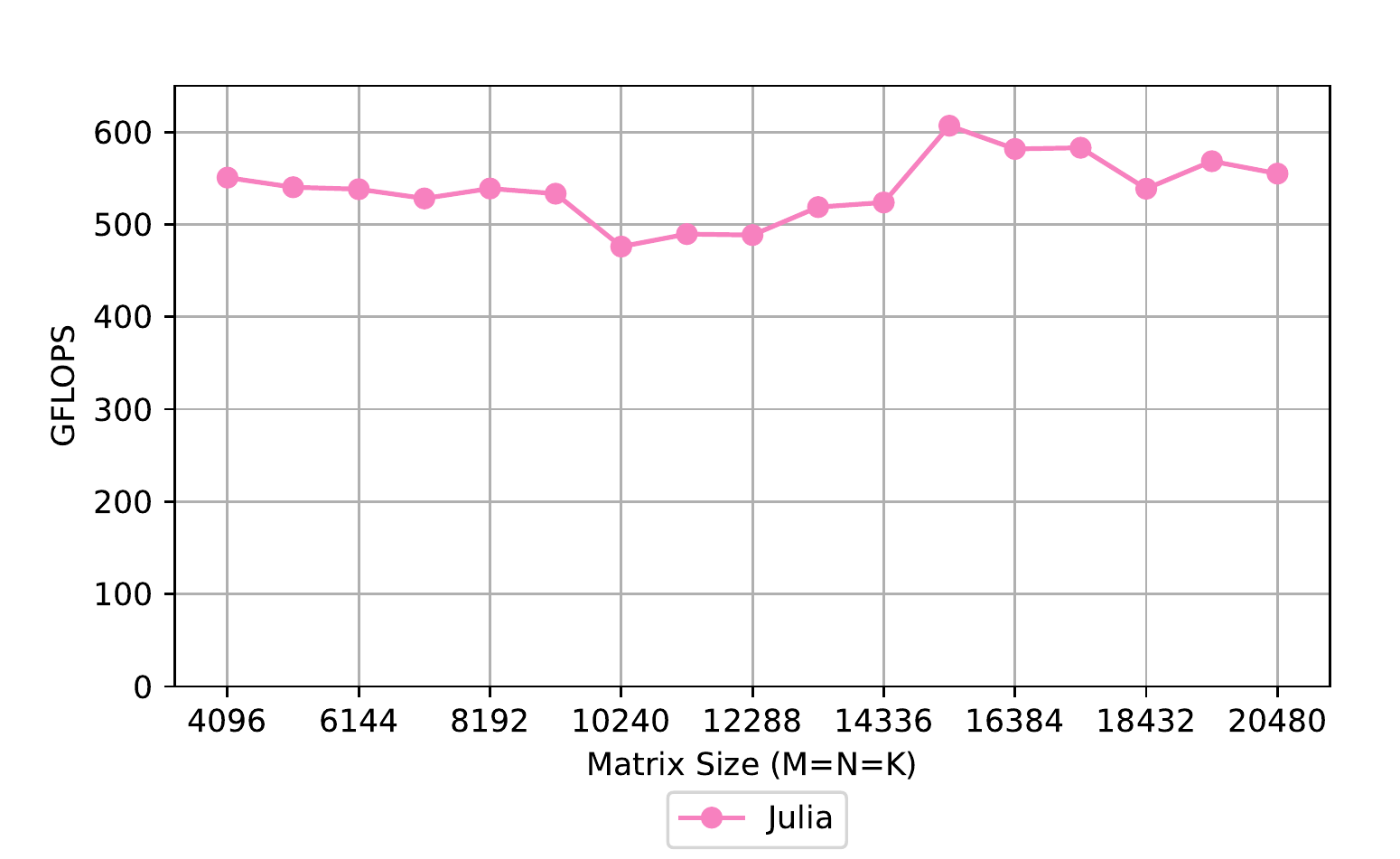}}
\caption{Wombat multithreaded CPU performance using 80 threads.}
\label{fig:WombatARMCPU}
\end{figure}

\paragraph{Wombat Arm Altra Ampere} Performance results on Arm CPUs are shown in Fig.~\ref{fig:WombatARMCPU} for (a)~double and (b)~single precision. Notably, Kokkos, which is using the OpenMP back end, experiences a slowdown in both cases. Meanwhile, Julia's performance is almost on par with the vendor OpenMP implementations.

Half-precision floating point (FP16) is not supported for Python/Numba regions combined with numpy's \texttt{Float16} random number capabilities, so input matrices were populated with 1s. Half-precision support in Julia is under
active development. We obtained very low performance on Crusher AMD CPUs (not reported in this work), and this is expected to improve as Julia's native FP16 support matures.\footnote{\url{https://github.com/JuliaLang/julia/issues/45542}} The Julia threads implementation on Arm worked seamlessly and provided the expected levels of performance, as shown in Fig~\ref{fig:WombatARMCPU}c. The literature also contains a recent discussion of Julia's FP16 performance on Arm systems~\cite{9912702}.
 
\subsection{GPU Performance}

\begin{figure}[ht]
    \centering
    \subfloat[Double precision (FP64)]{\includegraphics[width=3.5in,height=2.2in, valign=t]{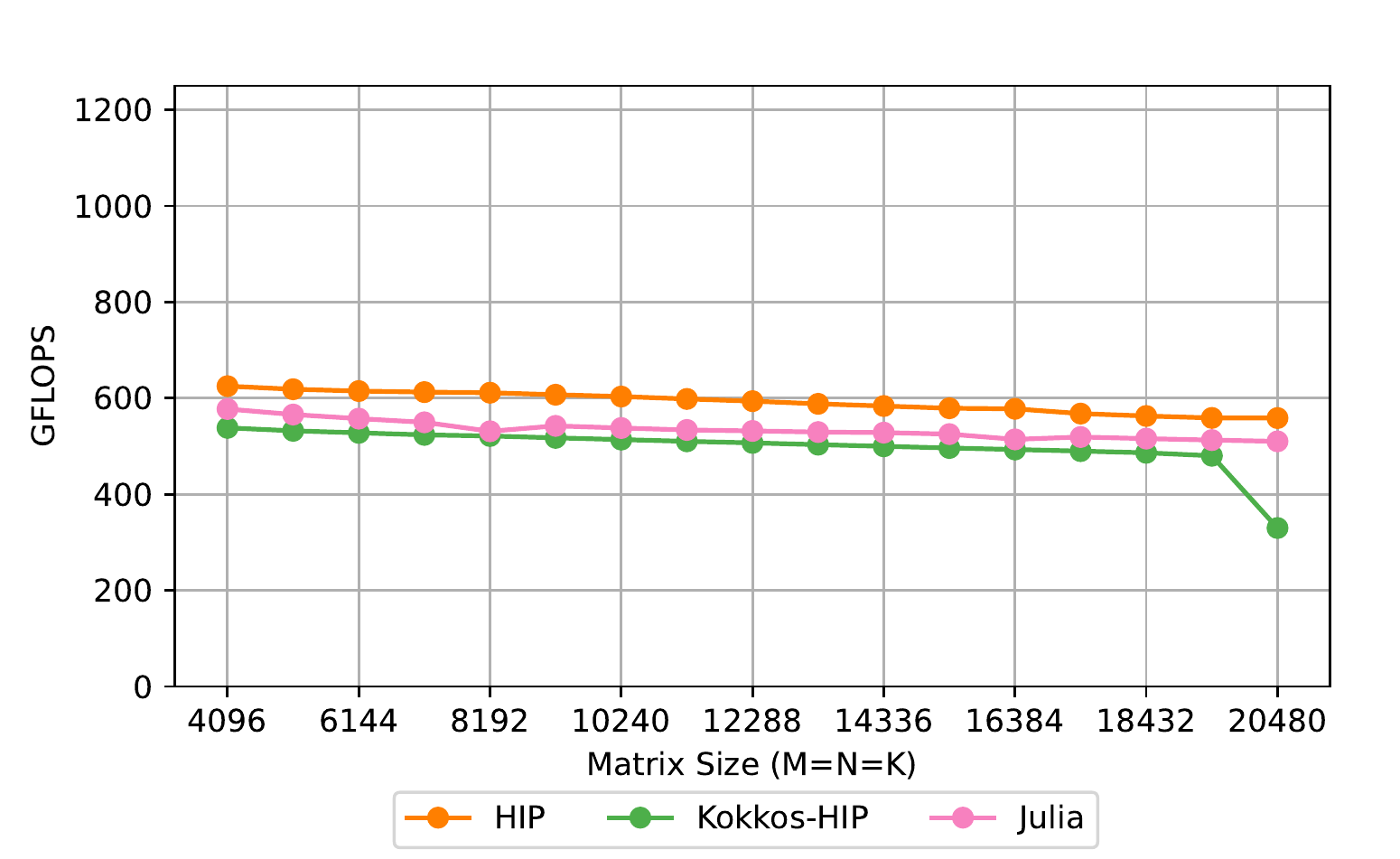}}
    \vfill
    \subfloat[Single precision (FP32)]{\includegraphics[width=3.5in,height=2.2in,valign=t]{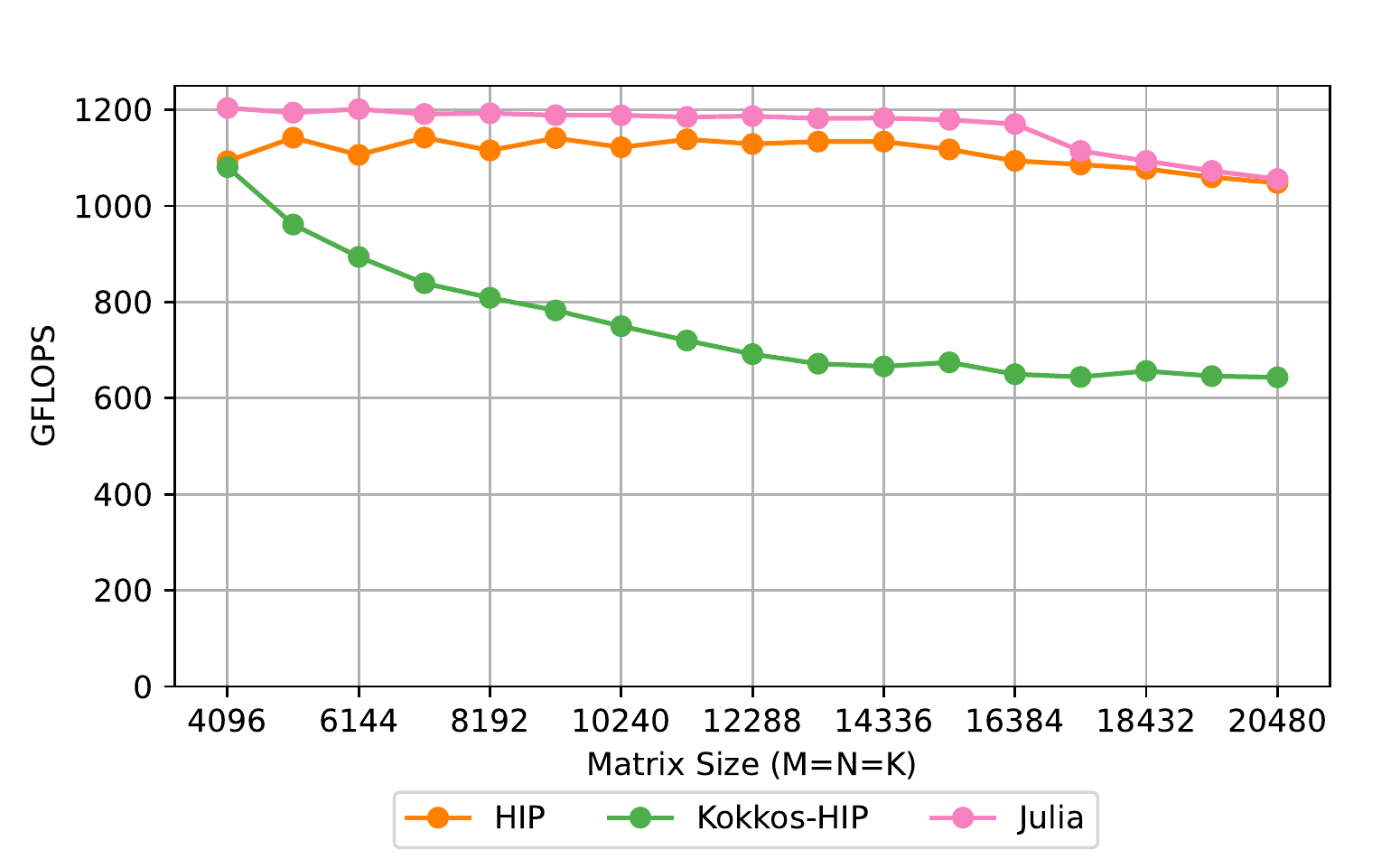}}
    \vfill
    \subfloat[Half precision (FP16)]{\includegraphics[width=3.5in,height=2.2in,valign=t]{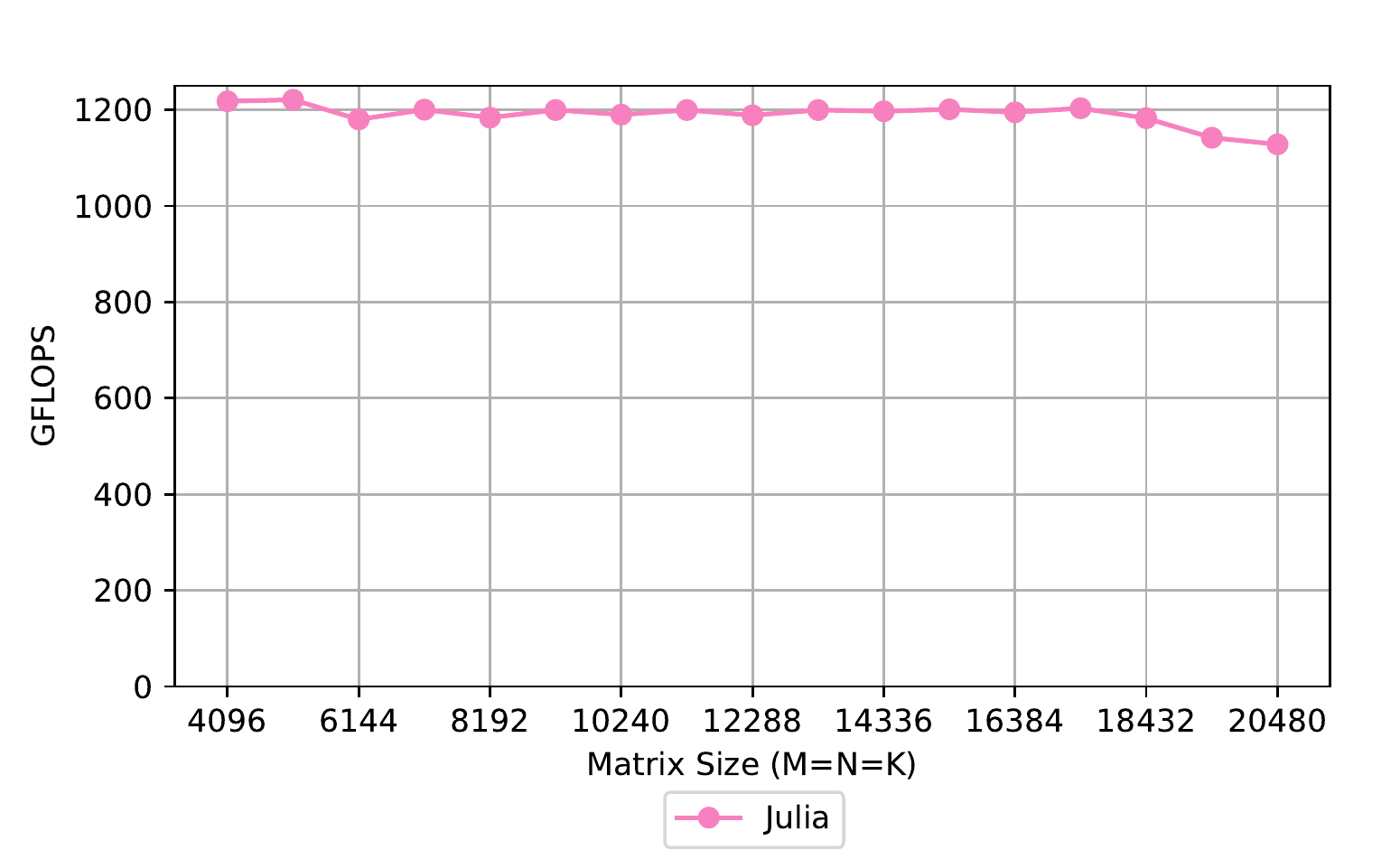}}
\caption{Simple GEMM performance on Crusher AMD MI250X GPU using $32\times32$ thread block sizes.}
\label{fig:CrusherAMDGPU}
\end{figure}

\paragraph{Crusher AMD MI250X} Figure~\ref{fig:CrusherAMDGPU} presents the simple GEMM performance for HIP, Kokkos-HIP, and Julia using AMDGPU.jl for different floating-point precisions on Crusher's AMD MI250X GPU, which is similar to the GPUs found in the OLCF's Frontier system (Table~\ref{tab:GPU}). Python/Numba is not supported on AMD GPUs. As shown in Fig.~\ref{fig:CrusherAMDGPU}a, for double-precision runs, the vendor-provided HIP implementation achieves the highest performance. This is followed by Julia using AMDGPU.jl and Kokkos/HIP, both of which reach competitive levels but still do not match HIP for all matrix sizes because the overheads introduced appear to be constant. Kokkos has a repeatable slowdown at the largest size, and this might require further investigation on this system. 
The performance for single-precision floating point is shown in Fig.~\ref{fig:CrusherAMDGPU}b. As expected, all models provide an increase in performance, but Kokkos~+~HIP exhibits a consistent decrease, which again requires further investigation. Interestingly, Julia with AMDGPU.jl shows slightly better performance than the vendor HIP implementation, although the differences become small for larger matrix sizes and this could simply be the variability on this particular system. Lastly, Julia AMDGPU.jl performance results are presented in Fig.~\ref{fig:CrusherAMDGPU}c for half-precision multiplications stored on a single-precision matrix (Fig.~\ref{fig:gemm}c). No noticeable improvements are shown when compared to single-precision runs. Nevertheless, other programming models do not provide seamless half-precision support, whereas Julia currently supports random number generation and kernel computations on AMD GPUs.

\begin{figure}[ht]
    \centering
    \subfloat[Double precision (FP64)]{\includegraphics[width=3.5in,height=2.2in, valign=t]{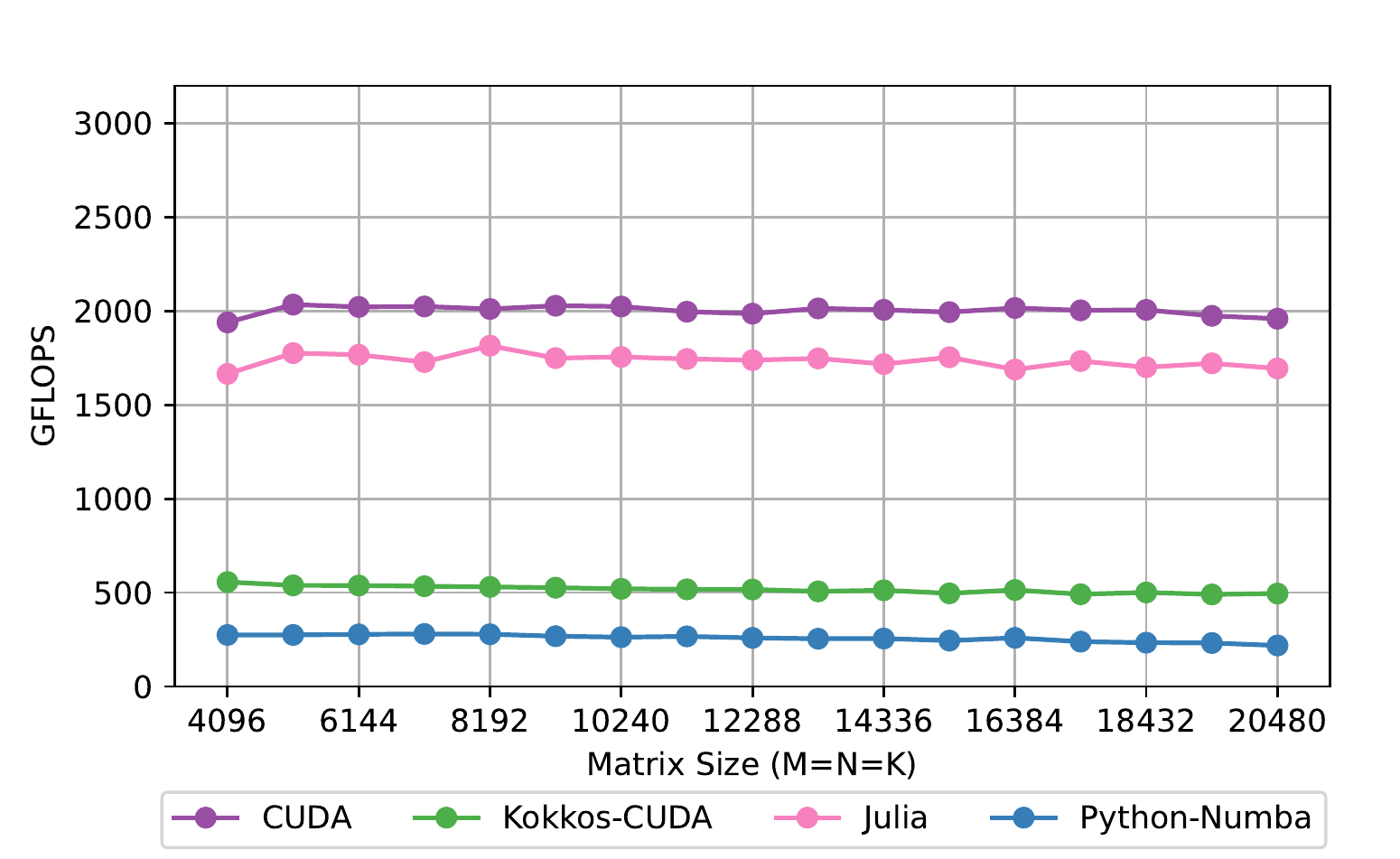}}
    \vfill
    \subfloat[Single precision (FP32)]{\includegraphics[width=3.5in,height=2.2in,valign=t]{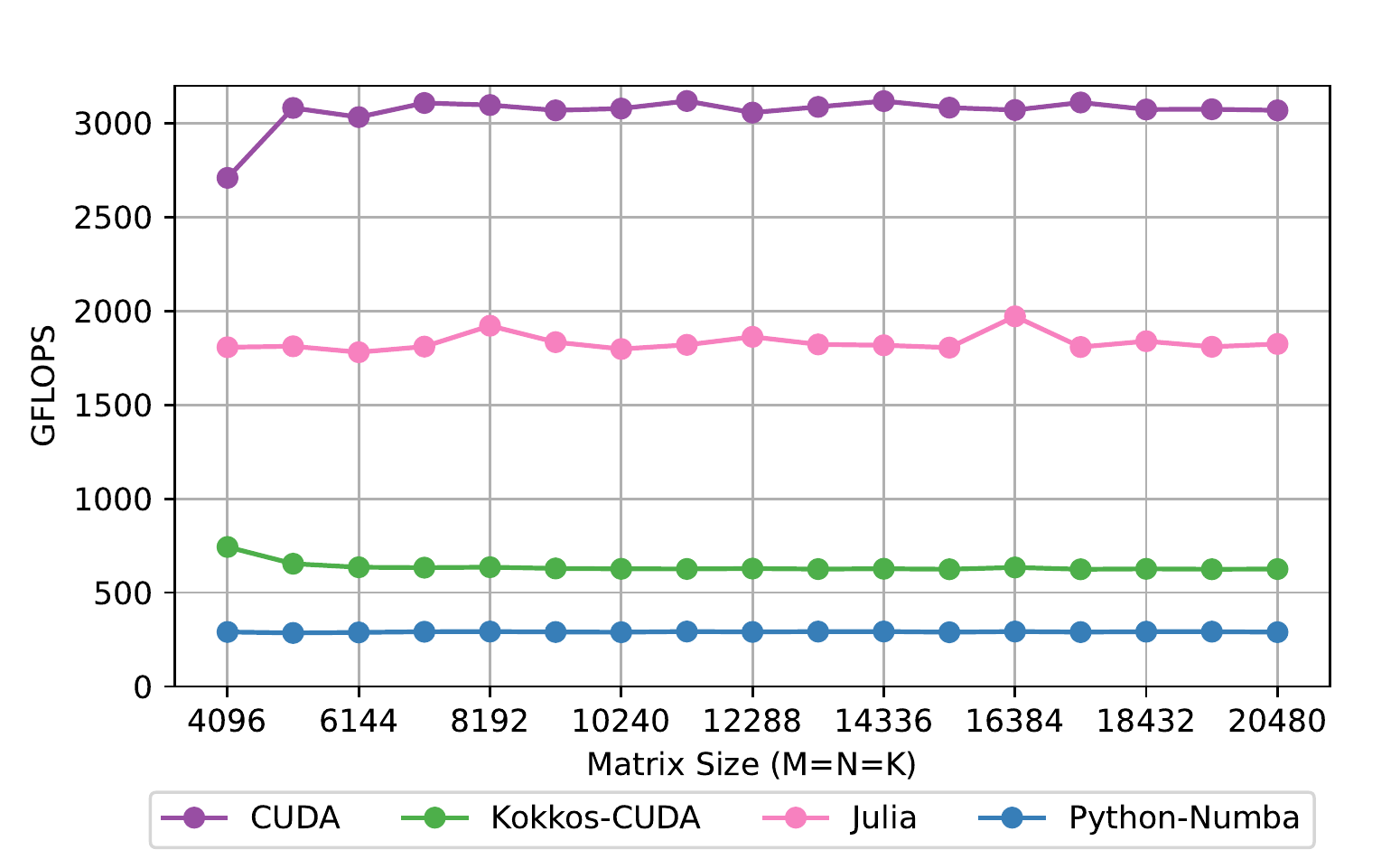}}
    \vfill
    \subfloat[Half precision (FP16)]{\includegraphics[width=3.5in,height=2.2in,valign=t]{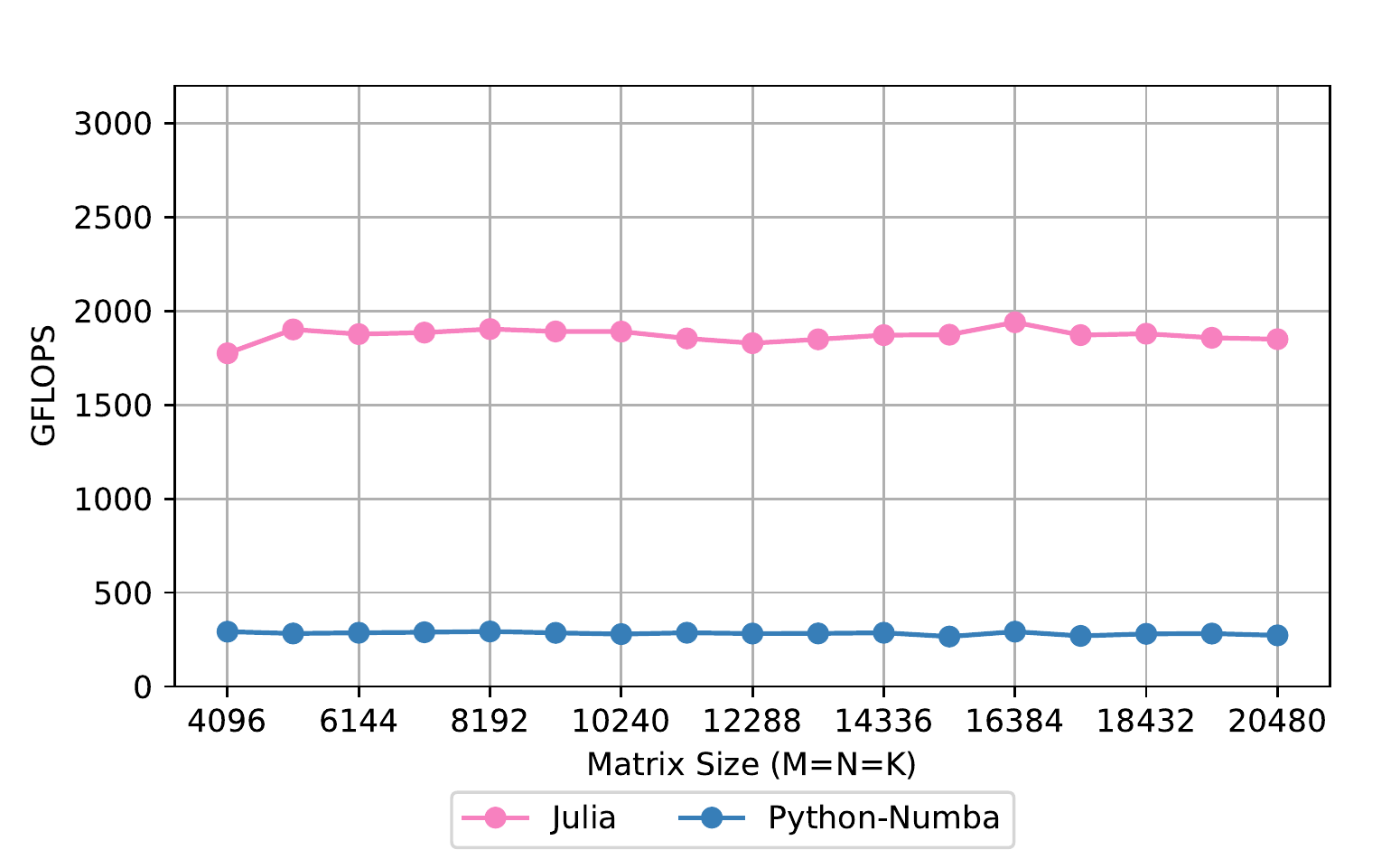}}
\caption{Simple GEMM performance on Wombat NVIDIA A100 using $32\times32$ thread block sizes.}
\label{fig:WombatNVIDIAGPU}
\end{figure}

\paragraph{Wombat NVIDIA A100} Figure~\ref{fig:WombatNVIDIAGPU} presents the simple GEMM performance for (1) CUDA, Kokkos/CUDA, and Julia using CUDA.jl and (2) Python/Numba using CUDA with different floating-point precisions on Wombat's NVIDIA A100 GPUs (Table~\ref{tab:GPU}). Double-precision runs shown in Fig.~\ref{fig:WombatNVIDIAGPU}a show that Julia using CUDA.jl has a constant overhead when compared to the vendor-provided CUDA implementation. The generated low-level Parallel Thread Execution (PTX) instruction set architecture (ISA), not shown here, indicated a difference in unrolled loop instructions, 2 for CUDA.jl and 4 in the native CUDA. Deeper investigation is required to generate more effective heuristic models for different kernel workloads.
Kokkos and Python/Numba using a CUDA back end consistently underperform, which raises questions about the configuration and/or actual GPU runs. Both Kokkos and Python/Numba were verified by using NVIDIA's \texttt{nvprof} profiler to corroborate GPU activity. Figure~\ref{fig:WombatNVIDIAGPU}b shows the performance for single-precision runs. As expected, the performance of the vendor-provided CUDA implementation increases significantly, whereas other implementations still present gaps for this case. Julia, Kokkos/CUDA, and Python/Numba show small performance increases of around 10\% between double- and single-precision runs.
Lastly, we show the half-precision results for the supported Julia with CUDA.jl and Python/Numba implementations. The Python half-precision implementation must be modified because random number generation is not supported using numpy's \texttt{float16} type. Nevertheless, we observed no performance gains over the single-precision counterparts.

%% file: 05-performance-portability.tex
Although, there is no agreed-upon metric for performance portability, we reference some of the proposals found in the current literature for parallel applications. However, the present work focuses on evaluating the performance portability of different programming models.

One of the first attempts to do so took place at the US Department of Energy's meeting on performance portability~\cite{doe-performance-portability}, during which different definitions were proposed. Pennycook et al.~\cite{PENNYCOOK2019947} proposed a unique definition, which has since been adopted by the HPC community: ``A measurement of an application’s performance efficiency for a given problem that can be executed correctly on all platforms in a given set.'' Since then, a set of different metrics and formulas was defined~\cite{MarowkaPDP21}. Recently, this definition was extended to define performance portability of a programming model as \textit{the ratio of the mean performance of a portable model and the mean performance of a non-portable one}~\cite{MarowkaHPCA22}.

The metric proposed to compute the performance portability of a programming model, $M$, was defined as follows:
\begin{equation}
    \Phi_M = \frac{\sum_{i \in T} e_i(a)}{\mid T \mid},
    \label{eq:PerformancePortability}
\end{equation}
where, in our case, $e_i(a)$ is the performance efficiency of the matrix-matrix multiplication of the portable programming model, $M$ (i.e., Kokkos, Julia, or Python-Numba), divided by the vendor-specific implementation.
For example, the performance efficiency of Julia on an MI250X AMD GPU would be formulated as follows:

\begin{equation}
    e_{MI250x} = \frac{Julia~Performance}{HIP~Performance}.
\end{equation}

The computed efficiencies for this simple kernel on each hardware target and the overall programming model, in Eq.~(\ref{eq:PerformancePortability}) are shown in Table~\ref{table:PerformanceEfficiency}. 
The vendor C/OpenMP performance was used as the architecture-specific reference model for CPU analysis, and CUDA and HIP performance was used as the architecture-specific reference model for NVIDIA and AMD GPUs, respectively.
The performance efficiency of Kokkos and Julia is similar, with the exception of $e_{A100}$. Python/Numba has the lowest numbers based on the performance results when considering that AMD GPUs are not supported. As for programming models efficiency for this simple kernel,
Julia has the best scores followed by Kokkos and Python/Numba. No significant difference is seen when comparing double-precision against single-precision analysis, and the
portability of all models is slightly lower for single-precision floating point computations.

\setlength\cellspacetoplimit{0.8ex}
\setlength\cellspacebottomlimit{0.8ex}
\begin{table}[ht]
\centering
\caption{Performance Efficiency of Kokkos, Julia, and Python/Numba for each experiment.}
\begin{tabular}{|| Sc | Sc | Sc | Sc ||}
\hline
 Architecture & Kokkos & Julia & Python/Numba\\
 \hline \hline
\hline
 \multicolumn{4}{||Sc||}{\textbf{Double precision}} \\
 \hline
 $e_{Epyc~7A53}$ & {\bf 0.994} & 0.912 & 0.550 \\
 $e_{Ampere~Altra}$ & 0.854 & {\bf 0.907} & 0.713 \\
 $e_{MI250x}$ & 0.842 & {\bf 0.903} & - \\
 $e_{A100}$ & 0.260 & {\bf 0.867} & 0.130\\
 \hline
 $\Phi_M$ & 0.738 & {\bf 0.897} & 0.348 \\
 \hline
 \multicolumn{4}{||Sc||}{\textbf{Single precision}} \\ 
 \hline
 $e_{Epyc~7A53}$ & {\bf 1.014} & 0.976 & 0.655 \\
 $e_{Ampere~Altra}$ & 0.836 & {\bf 0.900} & 0.400 \\
 $e_{MI250x}$ & 0.677 & {\bf 1.050} & - \\
 $e_{A100}$ & 0.208 & {\bf 0.600} & 0.095\\
 \hline
 $\Phi_M$ & 0.684 & {\bf 0.882} & 0.288 \\
 \hline
\end{tabular}
\label{table:PerformanceEfficiency}
\end{table}

%% file: 06-conclusions.tex
We studied the high-productivity, dynamic languages Julia and Python/Numba as high-level interfaces to LLVM and compared them with portable implementations of C/OpenMP and Kokkos.
Performance results and efficiency metrics for a simple hand-rolled gense matrix multiplication are presented on exascale node architectures---Wombat, which uses Arm Ampere CPUs and 2 NVIDIA A100 GPUs, and Crusher (Frontier's test bed), which is equipped with AMD EPYC 7A53 CPUs and 8 MI250X GPUs. 
Results for double- and single-precision floating point operations indicate that the default Julia implementations have comparable performance on these platforms. For CPUs, Julia performance was comparable to C/OpenMP combined with LLVM-based ArmClang and AMDClang vendor compilers.
For the AMD GPUs, Julia AMDGPU.jl performance was comparable to HIP.
Julia's productivity and performance benefits are the result of being designed from the ground up to leverage LLVM.
Nevertheless, there is still a performance gap on NVIDIA A100 GPUs for single-precision floating point cases and further investigation is needed into the default low-level PTX ISA code generated.
We observe that Python/Numba implementations still lack the support needed to reach comparable CPU and GPU performance on these systems, and AMD GPU support is deprecated. Kokkos provides an interesting approach for performance portability, which still depends on the back end and several compilation and policy settings.
Overall, Julia and Python/Numba programming models provide high-productivity CPU and GPU APIs for easy access to write LLVM-compiled kernels. Additionally, their powerful ecosystems for data analysis and workflows in HPC, seamless half-precision floating point support, and interoperability with vendor back ends all add value to the scientific discovery process. Future work should continue to explore their use in more complex HPC workloads as their LLVM-based implementations and supportive ecosystems become more mature to achieve desired performance portability on heterogeneous hardware.

%% file: a1-appendix.tex
The code used for this study is hosted on GitHub: \url{https://github.com/williamfgc/simple-gemm}. Each implementation has its own directory: C, Kokkos, Julia, and Python. The \texttt{scripts} directory contains the configurations for each experiment on OLCF systems. Figures~\ref{Wombat-CPU-C/OpenMP-script} and~\ref{Wombat-Julia-GPU-scripts} show examples of scripts to run C/OpenMP and Julia experiments on Wombat.

\begin{figure}[htp]
\centering 
\begin{minipage}{1\columnwidth}
\begin{minted}[breaklines, breakafter=d, fontsize=\scriptsize, frame=single, encoding=utf8]{bash}
#!/bin/bash
  
EXECUTABLE=../simple-gemm/c/gemm-dense-openmp64-armclang
module load ARM_Compiler_For_HPC/22.0
Ms=( 4096 5120 ... 19456 20480 )

export OMP_PROC_BIND=true
export OMP_PLACES=threads
export OMP_NUM_THREADS=80

for M in "${Ms[@]}"; do
  salloc -N 1 -p Ampere -t 10:00:00 srun -n 1 -c $t \
    $EXECUTABLE $M $M $M 5 > Ampere-ARMClang22-${t}t-${M}M_5s_threads.log 2>&1
done
\end{minted}
\end{minipage}
\caption{Wombat CPU C/OpenMP launch script.}
\vspace{-0.1in}
\label{Wombat-CPU-C/OpenMP-script}
\end{figure}

\begin{figure}[htp]
\centering 
\begin{minipage}{1\columnwidth}
\begin{minted}[breaklines, breakafter=d, fontsize=\scriptsize, frame=single, encoding=utf8]{bash}
#!/bin/bash

module load nvhpc-nompi/22.1
module load julia/1.7.3 
export JULIA_CUDA_USE_BINARYBUILDER=false
GemmDenseCUDADIR=../../simple-gemm/julia/GemmDenseCUDA
EXECUTABLE=$GemmDenseCUDADIR/gemm-dense-cuda.jl

Ms=( 4096 5120 ... 19456 20480 )

for M in "${Ms[@]}"; do

  salloc -N 1 -p Ampere -t 10:00:00 --gres=gpu:1 \ 
    srun -n 1 julia -O3 --project=$GemmDenseCUDADIR \
    $EXECUTABLE $M $M $M 5 \ 
    > A100-Julia1_7_3-${M}M_5s_F64.log 2>&1 &
done
\end{minted}
\end{minipage}
\caption{Wombat GPU Julia launch script.}
\vspace{-0.1in}
\label{Wombat-Julia-GPU-scripts}
\end{figure}

Tables~\ref{tab:CPU} and \ref{tab:GPU} describe the software stack and compilation flags used to generate all the experiments. Kokkos implementations are found in \texttt{simple-gemm/cpp/Kokkos/} with its own compilation framework.